\documentclass[prl,superscriptaddress,showpacs,twocolumn]{revtex4}
\usepackage{graphicx}

\begin{document}
\title{Distribution of reflection coefficients in absorbing chaotic microwave cavities}
\author{R.~A.~M\'endez-S\'anchez}
\affiliation{Centro de Ciencias F\'{\i}sicas UNAM, A.P 48-3, 62210, Cuernavaca, Morelos, M\'exico}
\author{U.~Kuhl}
\affiliation{Fachbereich Physik, Philipps-Universit\"at Marburg, Renthof 5, D-35032 Marburg, Germany}
\author{M.~Barth}
\affiliation{Fachbereich Physik, Philipps-Universit\"at Marburg, Renthof 5, D-35032 Marburg, Germany}
\author{C.~H.~Lewenkopf}
\affiliation{Instituto de F\'{\i}sica, UERJ, R. S\~ao Francisco Xavier 524, 20550-900 Rio de Janeiro, Brazil}
\author{H.-J.~St\"ockmann}
\affiliation{Fachbereich Physik, Philipps-Universit\"at Marburg, Renthof 5, D-35032 Marburg, Germany}
\date{\today}

\begin{abstract}
The distribution of reflection coefficients $P(R)$ for chaotic
microwave cavities with time-reversal symmetry is investigated in
different absorption and antenna coupling regimes.
For all regimes the agreement between experimental
distributions and random-matrix theory predictions is very good,
provided both the antenna coupling
$T_a$ and the wall absorption strength $T_w$ are
taken into account in an appropriate way.
These parameters are determined by independent
experimental quantities.
\end{abstract}


\maketitle

Wave scattering by chaotic and weak disordered systems
has motivated a rather intensive research activity.
The subject is common to several areas of physics, ranging from
nuclear, atomic, molecular, and mesoscopic physics
to classical wave scattering, like microwaves, sound, and light
(see Refs.~\cite{Bee97,Stoe99} for a review).
The striking feature shared by all such systems, provided they have 
a chaotic underlying classical dynamics, is that they show universal 
transmission fluctuations. These fluctuations are successfully described
by random-matrix theory \cite{Bro81,Guh98}.
Very recently a comprehensive treatment of absorption, ubiquitous in
experiments, was developed for systems without time-reversal symmetry (TRI) 
\cite{Sav,Fyo03}.
For systems with TRI, however, there is a rigorous theory only in 
limiting cases \cite{Sav}.


An analytical expression for the distribution $P(R)$ of the reflection
coefficient $R$ in the presence of absorption was derived
for an arbitrary number of open channels in the weak absorption limit
\cite{Bee01b}.
In the strong absorption limit $P(R)$ reduces to a simple exponential
\cite{Kog00}. In these two works perfect coupling
between the channels and the scattering region is assumed.
Furthermore, non-resonant backscattering processes \cite{Bro94,Bro95b,Gop98},
which are present in most experimental situations,
have been until now often neglected in cases that absorption is taken 
into account.
It is noteworthy that only few of these works are experimental
\cite{Dor90,Sto97,Sch01d}.

The purpose of this letter is to present experimental evidence that
random-matrix theory provides a quantitative understanding of the
universal reflection fluctuations in chaotic systems.
More specifically, we show that this can only be achieved if both the
coupling, so far mostly overlooked, and absorption strengths are properly
taken into account.
To this end we measure $P(R)$ in microwave cavities from
weak to strong absorption regimes.
A quantitative agreement between experiment and theory is observed
for all the cases we have studied. This is remarkable since all
theoretical parameters are directly obtained from averaged experimental
quantities and not from a fit of $P(R)$ to the data.

\begin{figure}
\mbox{\includegraphics[width=4.5cm]{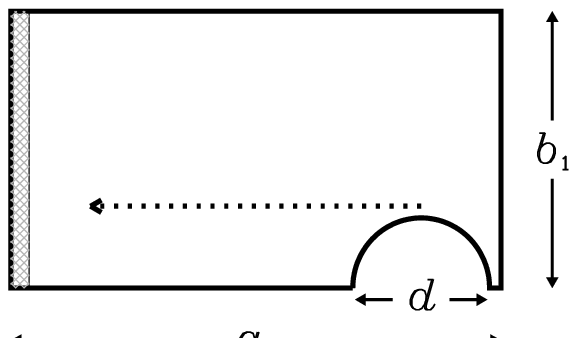} 
\hspace*{.2cm}
\includegraphics[width=4.5cm]{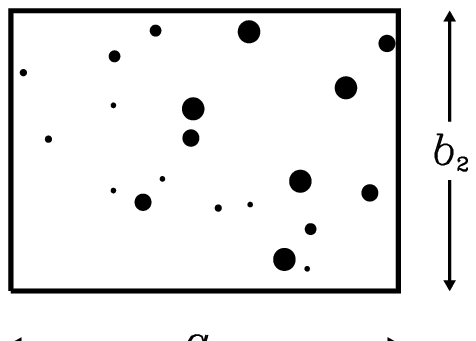}}    
\caption{
Sketch of the microwave billiards. (left) Half Sinai billiard
($a_1 = 43$ cm, $b_1 = 23.7$\,cm, height $h_1 = 7.8$\,mm).
The half disk ($d = 12$\,cm) was moved along $a_1$.
For the second set of measurements one wall was coated with an
absorber.
(right) Disordered billiard ($a_2 = 34$\,cm, $b_2 = 24$\,cm, height
$h_2 = 8$\,mm) \cite{Stoe01b}. The billiards are drawn at scale.}
\label{fig:cavities}
\end{figure}

\begin{figure}
\includegraphics[width=8.0cm]{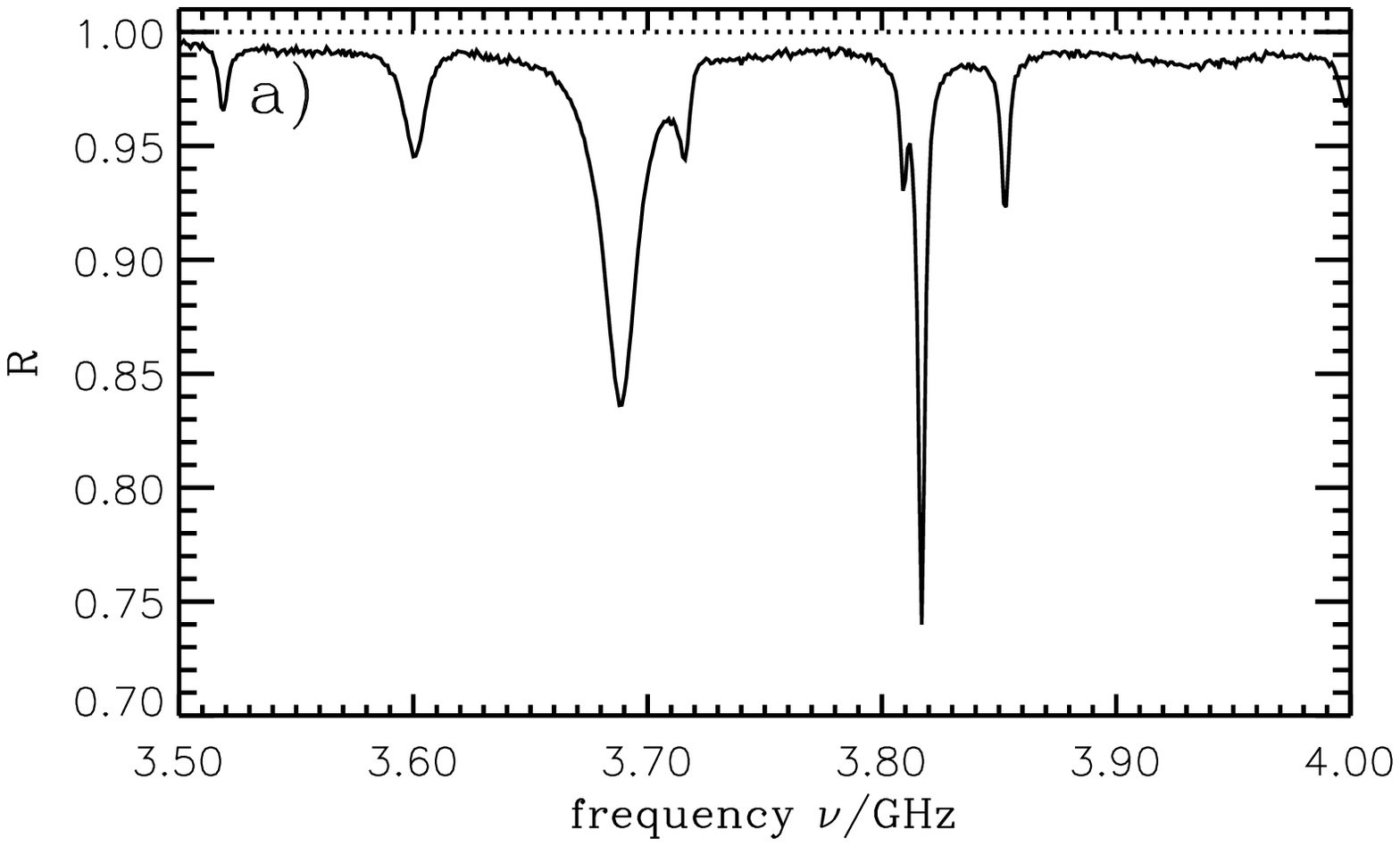} \\
{\includegraphics[width=8.0cm]{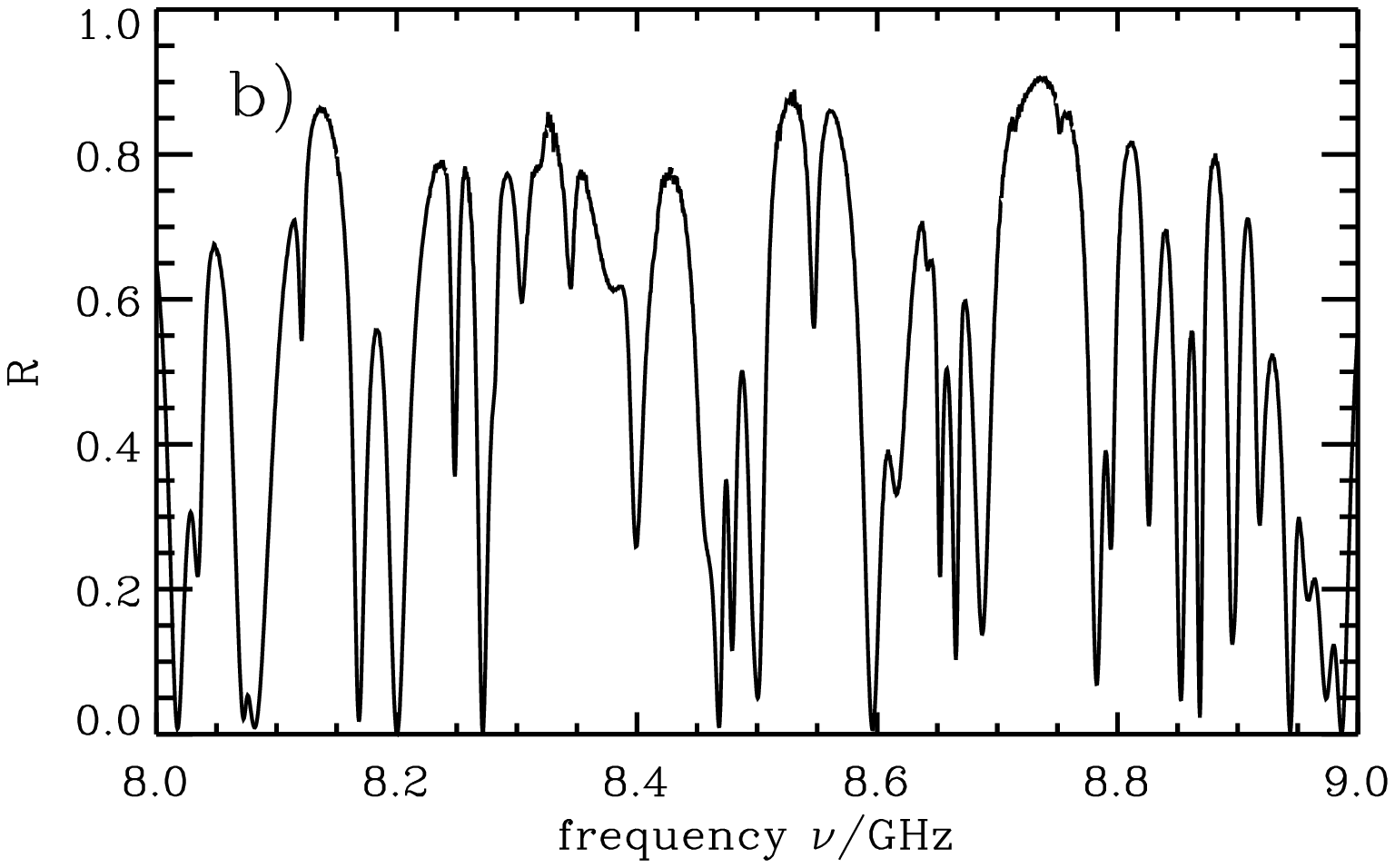}} \\
{\includegraphics[width=8.0cm]{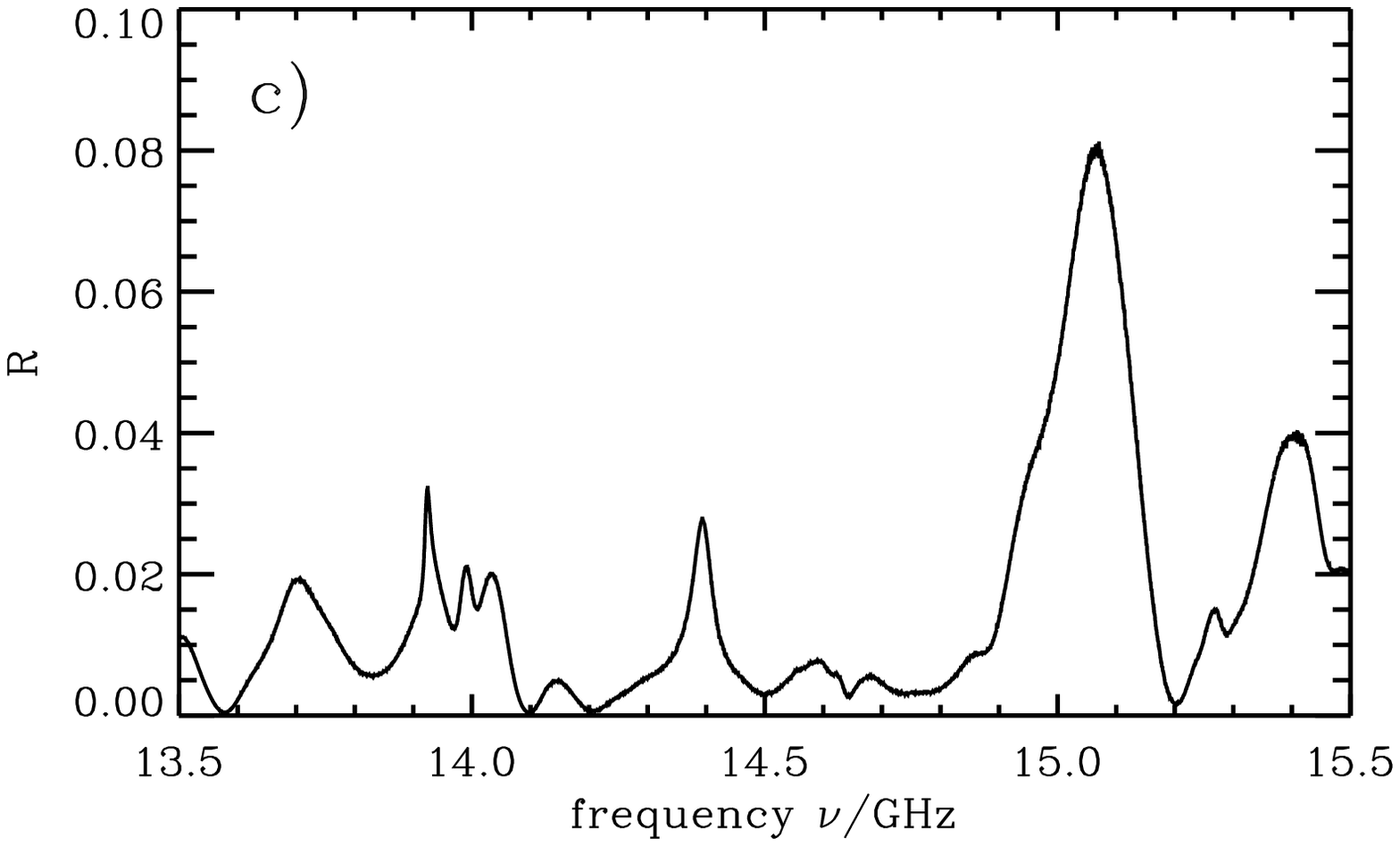}} \\
\caption{\label{fig:spectrum}
Typical absorption spectra for different absorption regimes:
(a) Weak absorption and weak coupling - disordered billiard,
(b) intermediate absorption and coupling - half Sinai billiard, and
(c) strong absorption and perfect coupling - half Sinai billiard
with absorber.}
\end{figure}

To study the dependence of $P(R)$ on absorption and coupling strengths,
we used three different flat microwaves cavities
and measured the reflection.
For the half Sinai billiard, Fig.~\ref{fig:cavities}(left),
57 reflection spectra were measured by sliding the half circular
inset along the wall over 28.5~cm in steps of 0.5~cm.
This is a practical way to improve the statistics,
notwithstanding the correlations among spectra.
For all cases only one antenna is coupled to the cavity.
For the second measurement, one wall of the
half Sinai billiard was coated with an absorber \cite{abs}.
The third measurement was performed
on a disordered billiard \cite{Stoe01b}, where 20 brass cylinders of 5
different sizes have been located at random within the billiard
(Fig.~\ref{fig:cavities}(right)).
The data were collected on a 5\,mm grid with 696 antenna positions.
We do not consider antenna positions closer than 15~mm
to the billiard walls, so only 533 measurements are included in the
statistical analysis.
The reflection coefficients $R$ were measured with vector network analyzers.
Typical absorption spectra for the three cases are shown in
Fig.~\ref{fig:spectrum}.

\begin{figure}
\includegraphics[width=8.0cm]{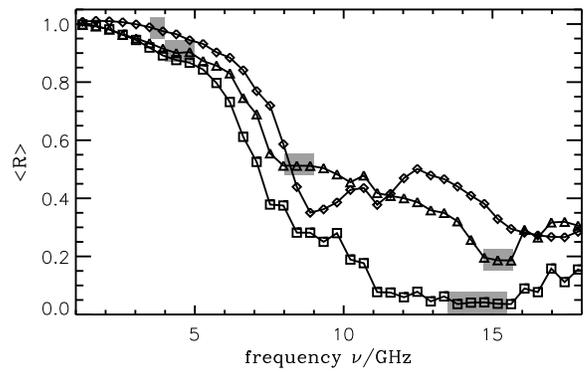}
\caption{Average reflection $\langle R \rangle$ as a function of
the frequency $\nu$ for: half Sinai (triangles), half
Sinai with absorber (squares) and disordered (diamonds), cavities.
The shaded boxes mark the cases subjected to a detailed
analysis. The frequency average is $\Delta \nu$=450 MHz.}
\label{fig:raverage}
\end{figure}

Figure~\ref{fig:raverage} shows the mean reflection coefficient
$\langle R \rangle$ as a function of the frequency $\nu$ for the
different cavities.
$\langle \cdots \rangle$ denotes averaging over a frequency range
$\Delta \nu$ (see Fig.\ \ref{fig:raverage} and over different cavity
geometries (half Sinai billiard) or different antenna positions (disordered
billiard).
Five representative windows,
where $\langle R \rangle$ is approximately constant,
are chosen for a detailed statistical investigation.
The windows correspond to the
shaded areas in Fig.~\ref{fig:raverage}.

We shall now discuss how $P(R)$ is obtained using random-matrix
theory \cite{Bee97}.
The scattering matrix $S$ is unitary and has a dimension $N$,
the number of open channels.
Since absorption opens new channels, the experimentally
accessible $N_{\rm exp}$-channel sector of $S$, which we call
$\widetilde{S}$, becomes subunitary. As a result,
the eigenvalues $R_1, R_2, \dots, R_{N_{\rm exp}}$
of $\widetilde{S}\widetilde{S}^\dagger$
lie between $0$ and $1$ \cite{Bee01b}.
In the whole frequency range the diameter of the antenna is small
compared to the wavelength implying a single coupled channel,
i.\,e., there is a single measurable reflection coefficient $R$.
The resonant $S$ matrix for quantum systems
\cite{Ver85a} can be used to describe cavity microwave scattering
\cite{Lew92,Fyo97}, and reads
\begin{equation}
\label{eq:S-Heidelberg}
S (E) = 1 - 2\pi i W^\dagger (E - H + i \pi W W^\dagger)^{-1} W \;.
\end{equation}
The quantum mechanical energy $E$ is related to the electromagnetic
frequency $\nu$ by Weyl's formula. In the high-frequency limit
$E\sim \nu^2$, whereas at low frequencies the dispersion relation
depends on the cavity shape.
$H$ describes the cavity quasi-bound spectrum, and $W$ contains the
coupling
matrix elements between the resonances and the scattering channels.
These are the absorption channels in the cavity
$N_w \gg 1$ and the microwave antenna $N_a=1$ \cite{Lew92}.
These absorption channels act as additional ``fictitious"
waveguides attached to the cavity \cite{comment}.

In our statistical treatment we take the matrix $H$ as being a member
of the Gaussian orthogonal ensemble, as usual for time-reversal
invariant chaotic systems.
Since the $H$ matrix is statistically invariant under orthogonal
transformations, the statistical properties of $S$ depend only
on the mean resonance spacing $\Delta$ (determined by $H$) and on
the traces of $W^\dagger W$.
The average scattering properties are best characterized by the
transmission coefficients of the antenna $t_a$ and of the fictitious
waveguides $t_w$. These coefficients are defined as
$t_c = 1-|\overline{ S_{cc}}|^2$, where $c$ is an arbitrary
channel and $\overline{\cdots}$ stands for an ensemble average.
The ergodic hypothesis identifies the later with the experimental
frequency averages.
Weak coupling corresponds to $t_c \rightarrow 0$,
or direct reflection from channel $c$ (no flux entering the cavity),
whereas $t_c \rightarrow 1$ is the limit of perfect coupling.
Hence, our model has two parameters,
namely, the antenna coupling $T_a = t_a$ and
the absorption strength $T_w = N_w t_w$.

Analytical results for $P(R)$ in the case of time-reversal invariant
systems are known only for the limiting cases of very weak ($T_w \ll 1$)
or strong absorption ($T_w \gg 1$) with perfect antenna coupling
($T_a = 1$). For most cases of interest one has to rely on numerical
simulations of the resonant $S$ matrix as given by
Eq.~(\ref{eq:S-Heidelberg}).
This is done as follows: first, $T_a$ and $T_w$ determine $W$
\cite{Ver85a}; second, by picking $H$ from the Gaussian orthogonal
ensemble we compute $S$.
In this study we generate $10^5$ realizations of $H$ with dimension $M=201$ 
and fix $N_w=200$ for each case of interest.
These values are justified as follows: The exact value of $N_w$ is not 
of relevance, provided $N_w \gg 1$, as demonstrated in Ref.~\cite{Sch03a}. 
The statistical theory requires $M\gg 1$, which is the case in our experiment.
Finally, for technical reasons $M>N_w$ \cite{Ver85a}.

\begin{table}
\begin{tabular}{|c|r|r|r|r|}\hline
billiard   & $\Delta \nu$/GHz
            & $\langle R \rangle$ & $T_a$ & $T_w$ \\ \hline
disordered     & 3.5- 4.0 & 0.981 & 0.011 &  0.49      \\ \hline
half Sinai     & 4.0- 5.0 & 0.906 & 0.116 &  0.56      \\ \hline
half Sinai     & 8.0- 9.0 & 0.517 & 0.754 &  2.42      \\ \hline
half Sinai     &14.7-15.7 & 0.185 & 0.989 &  8.40      \\ \hline
absorbing wall &13.5-15.5 & 0.039 & 0.998 & 48.00      \\ \hline
\end{tabular}
\caption{\label{tab:TaTw} Mean reflection coefficient $\langle R \rangle$,
antenna coupling $T_a$, and wall absorption $T_w$ for
the cases marked in Fig.\ \ref{fig:raverage}. $\Delta \nu$ is
the frequency range.}
\end{table}

The transmission coefficients $T_a$ and $T_w$ are obtained
directly from the experimental data as follows:
First, we use $t_a = 1-|\langle S_{aa} \rangle|^2$ to determine $T_a$
directly from the measured values of $S_{aa}(\nu)$. $T_w$ is then fixed by
adjusting the theoretical
mean reflection coefficient $\langle R \rangle$ to the experimental one.
In the low frequency regime, the values of
$\langle S_{aa}(\nu) \rangle$ and $\langle R \rangle$ cannot be extracted
correctly from the experimental data since the background of the
reflection spectrum varies slowly with $\nu$ and deviates from unity, as
shown in Fig.~\ref{fig:spectrum}(a). These variations of the background
cause a broadening of the distribution and a shift of
its maximum. This problem can be overcome as follows:
Since in this regime the resonances are nearly isolated, each of them
can be fitted by a complex Lorentzian. Using these fits a corrected
spectrum can be generated with a perfect background and suppression of noise.
From this spectrum $\langle S_{aa} \rangle$,
$\langle R \rangle$, and $P(R)$ are obtained.
The values for the mean reflection coefficient $\langle R \rangle$,
the antenna transmission $T_a$, and the wall absorption $T_w$
for the five regions under investigation are shown in Table~\ref{tab:TaTw}.

\begin{figure}
\includegraphics[width=8.0cm]{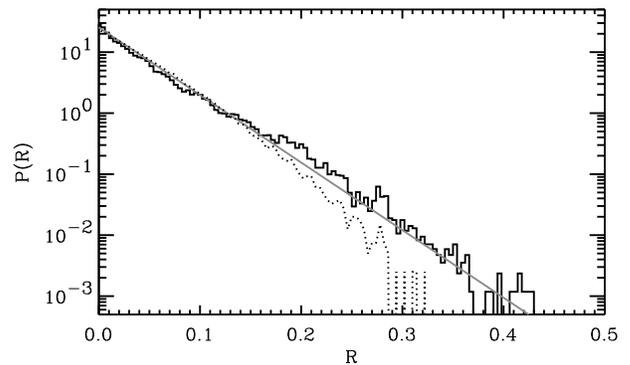}
\caption{Distribution of the reflection coefficient $R$ for the half Sinai
billiard in the strong absorption limit (13.5-15.5~GHz).
The solid line corresponds to Eq.~(\ref{eq:SAL}) and the dotted one to the
numerical simulation.}
\label{fig:pdrsinai}
\end{figure}

We first discuss the regime of strong absorption. The theoretical
distribution is given by \cite{Kog00}
\begin{equation}
\label{eq:SAL}
P(R)={\langle R \rangle}^{-1} {e^{-R/\langle R \rangle}},
\end{equation}
which no longer depends explicitly on the number of open channels.
Experimentally, the system with the largest absorption is the half Sinai
billiard with absorber as seen in Fig.~\ref{fig:raverage}.
A typical spectrum is presented in Fig.~\ref{fig:spectrum}(c).
The corresponding experimental distribution $P(R)$ is shown in
Fig.~\ref{fig:pdrsinai}, together with the prediction of
Eq.~(\ref{eq:SAL}) and the numerical simulation. The theory contains no
free parameters since the only inputs, namely $T_a$ and
$\langle R \rangle$, were independently extracted from the experimental
data.
A strong variation of $T_w$ in the investigated frequency range causes
the discrepancy between the experiment and the simulation.

\begin{figure}[tbp]
\includegraphics[width=8.0cm]{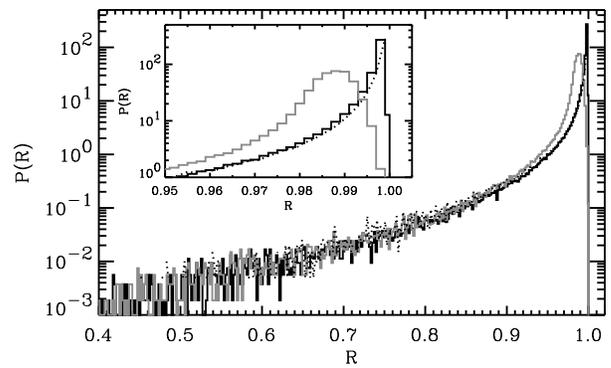}
\caption{Distribution of the reflection coefficients, $P(R)$, for weak coupling
strength and small absorption (3.5-4.0~GHz), obtained from the
Lorentzian fit corrected spectrum (for details see text).
For comparison the distribution obtained from the uncorrected spectrum is shown
in light grey.
The inset enlarges the region of $R$ between 0.95 and 1.005.
The dotted line is obtained from the random-matrix simulation.}
\label{fig:histoWAL}
\end{figure}

Let us now turn to the weak absorption regime, $T_w < 1$,
at low frequencies.
Unfortunately, the available analytical result \cite{Bee01b} cannot be used,
since it assumes perfect coupling, whereas in the present experiment,
the coupling is weak at low frequencies (see Table~\ref{tab:TaTw}).
A similar distribution for nearly perfect coupling has been
presented in Ref.~\cite{Dor90}, but no detailed analysis is performed.
In Fig.~\ref{fig:histoWAL} the distribution of $P(R)$ is shown for the
disordered cavity in the frequency region of 3.5 to 4.0 GHz, obtained
from the Lorentzian-fit corrected-spectra. This is beyond the
localization-delocalization transition observed at about 3~GHz
for the disordered cavity of Fig.\ \ref{fig:cavities}(right)
(a selection of eigenfunctions can be found in Ref.~\cite{Stoe01b}).
The results from the numerical simulation are plotted for
comparison with almost perfect agreement.

\begin{figure}
\includegraphics[width=8.0cm]{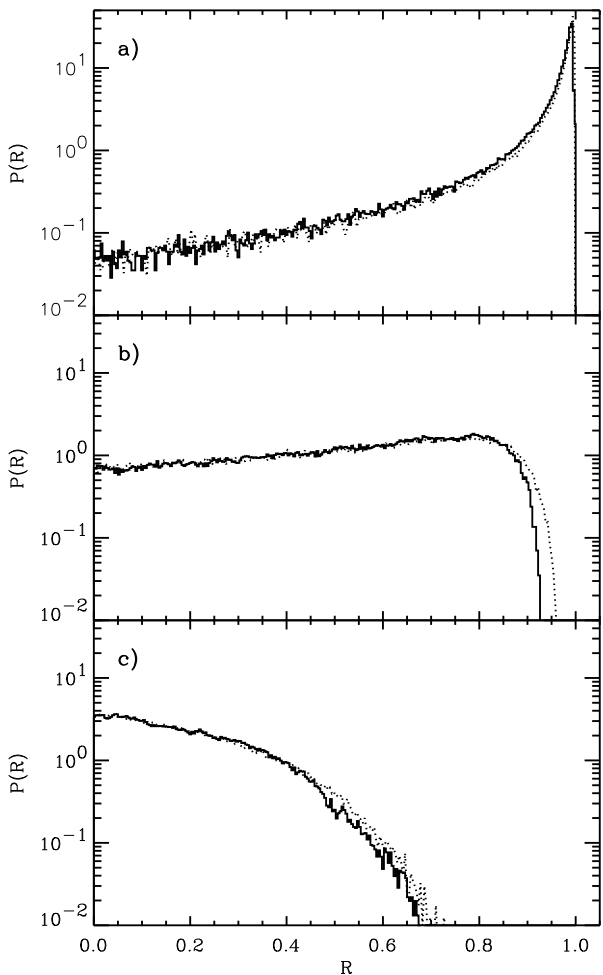}\vspace*{-2cm}
\caption{Distribution of the reflection coefficients $P(R)$ for the
intermediate absorption regime (half Sinai billiard).
The data were collected at different frequency windows
marked in Fig.~\ref{fig:raverage}.
The dashed lines stand for the random-matrix simulations.
In (a) the experimental distribution was obtained
form the Lorentzian fit corrected spectrum.
}
\label{fig:pdrsinaina}
\end{figure}

The experimental distributions $P(R)$ for intermediate absorption and
coupling regimes are presented in Fig.~\ref{fig:pdrsinaina}.
The data are taken from the half Sinai billiard depicted in
Fig.~\ref{fig:cavities}(left).
A typical spectrum for the intermediate regime is shown in Fig.~\ref{fig:spectrum}(b).
In all cases, a very good agreement between the experimental and theoretical $P(R)$ is
found.

It was essential to extract the coupling and
the absorption strength from two independent experimental quantities.
The coupling strength depends on the
phase variations of $S$ through $\langle S \rangle$,
whereas the absorption strength depends on the average
$\langle R \rangle$.
We obtain a good agreement between experiment and theory,
comparable to the one displayed in Fig.~\ref{fig:pdrsinaina},
by assuming perfect coupling ($T_a=1$) and thus overestimating
the absorption. Such agreement can lead to a wrong understanding
of the actual physical process at hand. This is an important
message to bear in mind when analyzing quantum mechanical
processes, where in most cases the phase of $S$ cannot be measured.

We have shown that microwave experiments with chaotic and disordered
billiards are excellent tools to investigate the universal distribution
of reflection coefficients with absorption. We studied a broad variety
of regimes, ranging from weak coupling and weak absorption to perfect
coupling and strong absorption. In all cases a very good agreement
between experiment and theory was found. This is particularly striking,
since the theoretical input parameters were fixed independently by
the experiment.

We thank C.~W.~J.~Beenakker, P.~A.~Mello, O.~Legrand, J. Flores, and
R.~Sch\"afer for helpful conversations and T.~H.~Seligman
and Centro Internacional de Ciencias (Cuernavaca) for hospitality.
This work was supported by DGAPA--UNAM, CONACyT (M\'exico),
CNPq (Brazil), and the DFG (Germany).

\bibliographystyle{prsty_long}

\bibliography{thesis,paperdef,paper,newpaper,book,pr_abso}

\begin{thebibliography}{10}
\bibitem{abs}
The absorber is a carbon loaded polyurethane foam of 5cm width
and an absorption larger than 20dB in the frequency range we have used.
\bibitem{comment}
The channels in Eq. (\ref{eq:S-Heidelberg}) are defined in the Hilbert space.
  Hence the ``fictitious" channels can, in principle, model wall as well as
  bulk absorption.
\end{thebibliography}

\end{document}